\documentclass[useAMS, usegraphicx, usenatbib]{mn2e}
\usepackage{aas_macros}
\usepackage{amsmath}
\usepackage{graphicx}
\usepackage{natbib}
\newcommand{\kunit}{\,h\,\mathrm{Mpc}^{-1}}
\newcommand{\runit}{\,h^{-1}\,\mathrm{Mpc}}
\newcommand{\runitk}{\,h^{-1}\,\mathrm{kpc}}
\newcommand{\munit}{\,h^{-1}\,\mathrm{M}_{\sun}}

\setcitestyle{authoryear,round,comma,aysep={},yysep={,},notesep={}}
\title[The effects of halo shape on clustering]{The effects of halo alignment and shape on the clustering of galaxies}
\author[M. P. van Daalen, R. E. Angulo and S. D. M. White]{Marcel P. van Daalen$^{1,2}$\thanks{E-mail: daalen@mpa-garching.mpg.de}, Raul E. Angulo$^{1}$ and Simon D. M. White$^{1}$\\
$^1$Max Planck Institute for Astrophysics, Karl-Schwarzschild Stra\ss{}e 1, 85741 Garching, Germany\\
$^2$Leiden Observatory, Leiden University, P.O. Box 9513, 2300 RA Leiden, The Netherlands}
\begin{document}
\date{Accepted 2012 May 31. Received 2012 May 21; in original form 2011 October 21}
\pagerange{\pageref{firstpage}--\pageref{lastpage}} \pubyear{2011}
\maketitle
\label{firstpage}
\begin{abstract}
We investigate the effects of halo shape and its alignment with larger scale structure on the galaxy correlation function. We base our analysis on the galaxy formation models of Guo et al., run on the Millennium Simulations. We quantify the importance of these effects by randomizing the angular positions of satellite galaxies within haloes, either coherently or individually, while keeping the distance to their respective central galaxies fixed. We find that the effect of disrupting the alignment with larger scale structure is a $\sim 2$ per cent decrease in the galaxy correlation function around $r \approx 1.8\runit$. We find that sphericalizing the ellipsoidal distributions of galaxies within haloes decreases the correlation function by up to $20$ per cent for $r \la 1\runit$ and increases it slightly at somewhat larger radii. Similar results apply to power spectra and redshift-space correlation functions. Models based on the Halo Occupation Distribution, which place galaxies spherically within haloes according to a mean radial profile, will therefore significantly underestimate the clustering on sub-Mpc scales. In addition, we find that halo assembly bias, in particular the dependence of clustering on halo shape, propagates to the clustering of galaxies. We predict that this aspect of assembly bias should be observable through the use of extensive group catalogues.
\end{abstract}

\section{Introduction}
Investigating how matter is organized in our Universe is one of the key ways in which we can test the validity of cosmological models and constrain their parameters. By comparing theoretical predictions to observed measures of structure, such as the galaxy correlation function or the matter power spectrum, one can reject some models and fine-tune others. It is, however, important to keep in mind the limitations of theoretical models, both numerical and analytical, when making this comparison, as these may limit the applicability of the results.

There are various ways in which one can predict the organization, or ``clustering'', of matter and galaxies theoretically. One can use fully hydrodynamical simulations, in which dark matter, gas and stars are treated explicitly, to follow the formation and evolution both of dark matter haloes and of the galaxies within them. For a recent review of the numerical methods behind such simulations, see \citet{Springel2010}. Such models are computationally expensive, limited to small volumes in comparison to recent galaxy surveys, and sensitive to the \textit{ad hoc} subgrid recipes required to include critical processes like star formation and feedback.

An alternative, first implemented by \citet{Kauffmann1999} \citep[see also][]{Springel2001, Springel2005}, is to combine N-body simulations of the growth of dark matter structures with semi-analytic models of galaxy formation (e.g. \citealp{WhiteFrenk1991, Kauffmann1993, Cole1994}; see \citealp{Baugh2006} for a review). A great advantage of semi-analytic simulations is that they require comparatively little CPU time even for a large underlying N-body simulation. This allows them to be run many times and on many haloes, so that one can explore the physical processes and the associated parameters that are required to produce galaxy populations in agreement with selected observational data (such as the galaxy stellar mass, luminosity or correlation functions). Such semi-analytic simulations do not focus on the properties of individual objects, but rather on the underlying statistical properties of the entire population. In this way, the relative importance of different physical processes can be examined as a function of the time and place where they are occurring. A disadvantage of such simulations is that they provide only very crude information on the structure of individual objects.

Yet another alternative is to take the statistical approach one step further. If one is interested only in the present-day clustering of galaxies, the physical processes associated with their formation and evolution may not be relevant. One can then populate the haloes in an N-body simulation with galaxies using a purely statistical model that depends on current halo properties, for example halo mass. Galaxy clustering can then be described in terms of the clustering of the haloes. This approach is known as halo occupation distribution modelling, or simply HOD modelling \citep[see][ for a review]{CooraySheth2002}. Typically, central and satellite galaxies are treated separately, as each halo will contain one and only one of the former but may contain none or many of the latter \citep{Kauffmann1999,Kravtsov2004,Zheng2005}. The satellite galaxies assigned to a halo are usually assumed to be spherically distributed following a standard profile such as that of \citet*{Navarro1997}. Attempts at including substructure or an environmental dependence have also been made \citep[e.g.][]{Giocoli2010,Gil-Marin2011}. Note that by assuming spherical symmetry some information is lost. As the simulations of \citet{Davis1985} first showed, cold dark matter haloes are typically strongly ellipsoidal. If the distribution of galaxies follows the mass distribution, this would leave an imprint on the galaxy correlation function on small scales. Furthermore, halo ellipticity may also have an effect on larger scales. If neighbouring haloes are aligned, as expected from tidal-torque theory, this will boost the correlation on scales corresponding to the typical separations between haloes.

The ellipticity and intrinsic alignment of dark matter haloes and their galaxy populations have been the subject of many earlier studies, e.g. \citet{CarterMetcalfe1980, Binggeli1982, West1989, Splinter1997, JingSuto2002} and \citet{Bailin2005}, and recently \citet{Paz2011} and \citet{Smargon2012}. Most relevant to the current work are the studies by \citet{SmithWatts2005} and \citet{Zu2008}. The former authors investigated the effects of halo triaxiality and alignment on the matter power spectrum in the halo model framework. Inspired by the results of simulations, they took a purely analytic approach in which they re-developed the halo model to account for ellipsoidal halo shapes. \citet{Zu2008}, on the other hand, used the semi-analytical models of \citet{DeLuciaBlaizot2007} to investigate environmental effects, including that of halo ellipticity, on the galaxy correlation function in both real and redshift space.

There exists a deeper connection between halo shape and clustering that we also explore in this paper. As \citet{Bett2007} and \citet{FaltenbacherWhite2010} have previously shown, at fixed mass the clustering of haloes depends on their shape. This is probably a reflection of assembly bias \citep[i.e. the dependence of halo clustering on properties other than mass,][]{GaoWhite2007}. More specifically, the most spherical haloes in their samples cluster significantly more strongly than average, and the most aspherical more weakly. The known correlations between formation time and halo shape \citep[e.g.][]{Allgood2006,Ragone-Figueroa2010}, and between formation time and clustering strength \citep[e.g.][]{Gao2005,Wechsler2006,Wetzel2007,Jing2007} point in this direction but may by themselves not be strong enough to explain the magnitude of the effect. Here, we investigate whether this shape-dependence is also seen in the clustering of the galaxies. If so, this would bring us one step closer to measuring assembly bias directly in observations. We are herein also motivated by the results from \citet{Zhu2006} and \citet{Croton2007}, who showed that assembly bias in general is indeed expected to propagate to galaxy clustering.

In this paper, we expand upon previous work by investigating the effects of alignment and ellipticity on the galaxy correlation function using the Millennium Simulation \citep{Springel2005} and the semi-analytic models of \citet{Guo2011}. In Section \ref{sec:methods}, we discuss these simulations and our methods for quantifying the effects of alignment and ellipticity. We also outline our procedure for determining the shape-dependence of galaxy bias. We show our results in Section \ref{sec:results}, and present our conclusions in Section \ref{sec:summary}.

\section{Methods}
\label{sec:methods}
\subsection{Simulation and SAM}
We make use of the galaxy catalogues generated by \citet[][, hereafter G11]{Guo2011}, who implemented galaxy formation models on the Millennium Simulations \citep{Springel2005,Boylan-Kolchin2009}. The Millennium Simulation (MS) is a very large cosmological N-body simulation in which $2160^3$ particles were traced from redshift $127$ to the present day in a periodic box of side $500\runit$, comoving. The Millennium-II Simulation (MS-II) follows the same number of particles in a box of side $100\runit$ and so has $125$ times better mass resolution. Both simulations assume a $\Lambda$CDM cosmology with parameters based on a combined analysis of the 2dFGRS \citep{Colless2001} and the first-year WMAP data \citep{Spergel2003}. These cosmological parameters, given by $\{\Omega_\mathrm{m}, \Omega_\mathrm{b}, \Omega_\mathrm{\Lambda}, \sigma_8, n_\mathrm{s}, h\}$ = $\{0.25, 0.045, 0.75, 0.9, 1.0, 0.73\}$, are not consistent with the latest analyses of the CMB data, for example the seven-year WMAP results \citep{Komatsu2011}. In particular, the more recent data prefer lower $\sigma_8$ and higher $\Omega_\mathrm{m}$ values.\footnote{See \citet{AnguloWhite2010} for a method to correct for this.} We will only make relative comparisons between clustering statistics here, and do not expect our results to be significantly influenced by these small parameter differences \citep[see][]{Guo2012}.

The galaxy formation models of G11 allow galaxies to grow at the potential minima of the evolving population of haloes and subhaloes in the simulations. Each Friends-of-Friends (FoF) group contains a central galaxy at the potential minimum of its main subhalo, and may contain many satellite galaxies at the centres of surrounding subhaloes. In some cases, due to tidal effects, a satellite galaxy may be stripped of its dark matter to the point where its subhalo is no longer identified as a bound substructure, turning the galaxy into an ``orphan''. Such galaxies follow the orbit of the dark matter particle that had the highest binding energy immediately before subhalo disruption, except that their distance to the central galaxy is artificially decreased until they merge with it in order to mimic the effects of dynamical friction. We note that the treatment of the orbits of orphans is approximate, and for example does not include the expected circularization of the orbits \citep[e.g.][]{Boylan-Kolchin2008}. The models also include treatments of star formation, gas cooling, gas stripping, metal enrichment, supernova and AGN feedback, and galaxy mergers. For more details about the SAM, as well as the treatment of different types of galaxies, we refer to G11. For our purposes, it is enough to note that the predicted clustering of galaxies is quite a close match to that seen in the Sloan Digital Sky Survey \citep{Guo2011}.

\subsection{Calculation of the galaxy correlation function}
The galaxy two-point correlation function, $\xi(r)$, measures the clustering of galaxies as a function of scale. It effectively encodes the excess probability of finding a pair of galaxies at a given separation $r$, relative to the expectation for a uniform random distribution. In what follows, we will be interested in scales $30\runitk < r < 50\runit$, as these are both well-resolved and well-sampled by the simulation. In order to get accurate results over this full range, we calculate the correlation function by direct pair counts on small scales (i.e. $r \la 4\runit$) and use an approximate but accurate method to calculate it on intermediate and large scales.

A direct calculation of this function scales as the number of galaxies squared and is thus unfeasible for the large sample analysed here. We therefore speed up the calculation by mapping galaxies onto a grid, and we correlate the mean density contrast in each grid cell with that of every other (a method previously employed by, for example, \citealp{Barriga2002}, \citealp{Eriksen2004} and \citealp{Sanchez2008}). We improved the performance on intermediate scales by folding the density field onto itself before its autocorrelation is calculated \citep[see e.g.][]{Jenkins1998}. We do not go into these methods here, but note that tests against higher-accuracy calculations show that the error in the ratio of the correlation functions, which is the relevant quantity for our main results, is less than $1$ per cent on all scales considered. For the correlation functions we calculate to determine the galaxy bias a direct pair count over the full range of scales \emph{is} feasible, as there we only consider relatively small subsets of galaxies (see \S\ref{subsec:biasresults}).

\subsection{Testing the importance of alignment and ellipticity}
During their lifetime, haloes merge and may accrete more subhaloes. The accretion of mass is not isotropic since matter flows in preferentially along filaments (see e.g. \citealp{Tormen1997,Colberg1999} or more recently \citealp{Vera-Ciro2011}). As a result, the distribution of subhaloes and thus galaxies within a FoF group is generally not isotropic either, but is instead approximately ellipsoidal, following the mass and aligning with surrounding large-scale structure \citep[see e.g.][]{Angulo2009}. To test whether alignment with neighbouring structure has an effect on clustering statistics, we randomly rotate the haloes around their centres and see if this systematically alters the galaxy correlation function. More precisely, we rotate the satellite population of each FoF group bodily around the central galaxy to a new randomly chosen orientation, and we repeat this process for every FoF group in the simulation. We stress that this transformation preserves the numbers, properties, and relative positions of the galaxies in every halo; only the orientations of the distributions change. We then calculate the galaxy correlation function for the new distribution, and compare it to the original. If alignment with large-scale structure is important, one would expect to see the correlation decrease systematically on scales slightly larger than individual haloes. To estimate the uncertainty in our results, we have repeated this process $25$ times, each time with a different set of randomly chosen angles.

The effect of halo ellipticity is tested in a similar way. Here, we randomly rotate the position of each individual satellite galaxy around its central, rather than rotating all satellites together. In this way, the galaxy distribution within each halo is sphericalized. Since the distribution of galaxies within haloes is typically ellipsoidal, this process should increase the average distance between galaxies, thus decreasing the correlations between galaxies in the same halo. We note that \citet{Zu2008} investigated the effect of halo ellipticity in the same way.

\subsection{Testing the dependence of galaxy bias on halo shape}
\label{subsec:bias}
\citet{FaltenbacherWhite2010} showed that the clustering of haloes depends on the shape of the halo, defined as $s=c/a$, where $c$ and $a$ are eigenvalues of the inertia tensor ($a>b>c$). Specifically, they showed that the large-scale bias of haloes with more spherical shapes is larger than average, while the inverse is true for the most aspherical haloes. They also found that this difference decreases with equivalent peak height, $\nu(M,z)=\delta_\mathrm{c}(z)/\sigma(M,z)$, where $\sigma(M,z)$ is the root-mean-square linear overdensity within a sphere which contains the mass $M$ in the mean, and $\delta_\mathrm{c}(z)$ is the linear overdensity threshold for collapse at redshift $z$. Here, we are interested in seeing if this shape-dependent clustering, which might reflect the assembly bias of the haloes, is also recovered from the galaxy distribution.

We use the halo shape data from \citet{FaltenbacherWhite2010}, who calculated the inertia tensor from the dark matter particles belonging to the most-pronounced subhalo\footnote{The most-pronounced subhalo may differ from the most massive subhalo only if the FoF groups hosts two or more subhaloes with roughly the same mass. In this case one of these is arbitrarily assigned to be the most massive. The most-pronounced subhalo is more consistently defined by using the halo merger tree. It is also the subhalo hosting the most luminous galaxy in the FoF group.} of each FoF halo, which on average comprises $\sim 80$ per cent of its mass. To ensure that the shapes were accurately determined, only haloes with at least 700 particles were considered, corresponding to a minimum (sub)halo mass $M=6.02\times 10^{11}\munit$. We compare this to the shape measured from the galaxy distribution in the same way, using all galaxies with stellar masses $M_*>10^9\munit$ within a sphere of radius $R_{200}$, defined as the radius enclosing 200 times the mean density of the Universe, centred on the central galaxy.  Using such a distance cut makes it easier to compare our results to observations -- in fact, a similar procedure is often followed when determining the richness of real groups and clusters. Rejecting galaxies outside the virial radius slightly biases us to measure more spherical shapes, but we have checked that this effect is small and does not significantly affect our results.

\begin{figure}
\begin{center}
\includegraphics[width=1.0\columnwidth, trim=7mm 8mm 0mm 8mm]{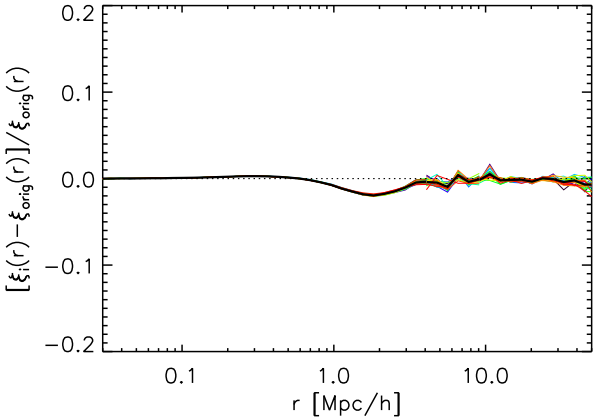}
\caption{The effect of halo alignment on the galaxy correlation function. The x-axis shows the real-space separation $r$, while the y-axis shows the fractional difference between the correlation function after random bodily rotations are applied, and that of the original, unrotated sample. All simulated galaxies with $M_*>10^9\munit$ from the $z=0$ catalogue of G11 have been used here. The bin size is roughly $0.07\,\mathrm{dex}$. Each of the 25 thin, coloured lines represents a different set of random rotations, and the thick, black line shows the average of these. There is a clear signal around $r \approx 1.8\runit$, where the correlation function is lowered by roughly $2$ per cent.}
\label{correl_rot1}
\end{center}
\end{figure}

After splitting the galaxies by the shape of their halo (measured either from the dark matter or from the galaxies themselves) we determine the large-scale galaxy bias factor $b_\mathrm{gal}$ for each subsample. Here, too, we follow \citet{FaltenbacherWhite2010}, who in turn followed the approach of \citet{GaoWhite2007}. The bias is computed as the relative normalization factor that minimizes the mean square of the difference $\log(\xi_\mathrm{gm})-\log(b_\mathrm{gal}\xi_\mathrm{mm})$ for four bins spaced equally in $\log r$ in the range $6<r<20\runit$. Here $\xi_\mathrm{mm}$ is the dark matter autocorrelation function and $\xi_\mathrm{gm}$ is the cross-correlation function of galaxies and dark matter. Note that unlike \citet{FaltenbacherWhite2010} we are only interested in the results at $z=0$.

\section{Results}
\label{sec:results}
\subsection{Alignment and ellipticity}
\label{sec:rot}
We will first discuss our results for ``bodily'' rotations, which test the effect of halo alignment. Figure~\ref{correl_rot1} shows the fractional difference between the correlation functions of the ``rotated'' and original samples, plotted against the real-space separation $r$. We have only used those galaxies from the catalogue generated by G11 that have a stellar mass $M_*>10^9\munit$, as the Millennium Simulation is not complete below this limit. This provides a sample of $5\,200\,801$ galaxies. We note that increasing this mass limit by a factor of ten does not influence our results, either qualitatively or quantitatively. All random rotations are applied prior to the mass cut in order to avoid problems in cases where a central galaxy below the limiting mass has satellites above it. Coloured lines indicate different sets of rotations, while the thick, black line shows the average of these. There is clearly a significant dip around $r \approx 2\runit$, with a depth of $2$ per cent. This is due to the disruption of the alignment between haloes and surrounding structure. Note that the scatter is extremely low, due to the large number of objects (in fact, the uncertainty at large scales is dominated by the errors due to our approximate calculation of the correlation). Neglecting the orientation of haloes when populating them with galaxies will therefore have a modest, but significant, effect on the derived correlation function.

\begin{figure}
\begin{center}
\includegraphics[width=1.0\columnwidth, trim=7mm 8mm 0mm 8mm]{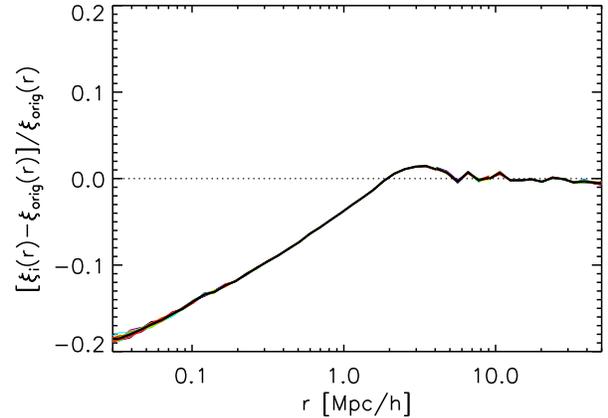}
\caption{The effect of halo ellipticity on the galaxy correlation function. The bin size, axes and lines are as in Figure~\ref{correl_rot1}, although the bodily rotations have been replaced by random independent rotations of satellites around their central galaxies. A peak of $1-2$ per cent can be seen around $r \approx 3.5\runit$, but the largest effect is seen on small scales, where the correlation function is systematically lowered by up to $\sim 20$ per cent. Note also that the scatter has been greatly reduced relative to Figure~\ref{correl_rot1}. This is mainly due to the larger number of random rotations used when rotating satellites separately.}
\label{correl_rot2}
\end{center}
\end{figure}

\begin{figure*}
\begin{center}
\begin{tabular}{ccc}
\includegraphics[width=1.0\columnwidth, trim=7mm 8mm 0mm 8mm]{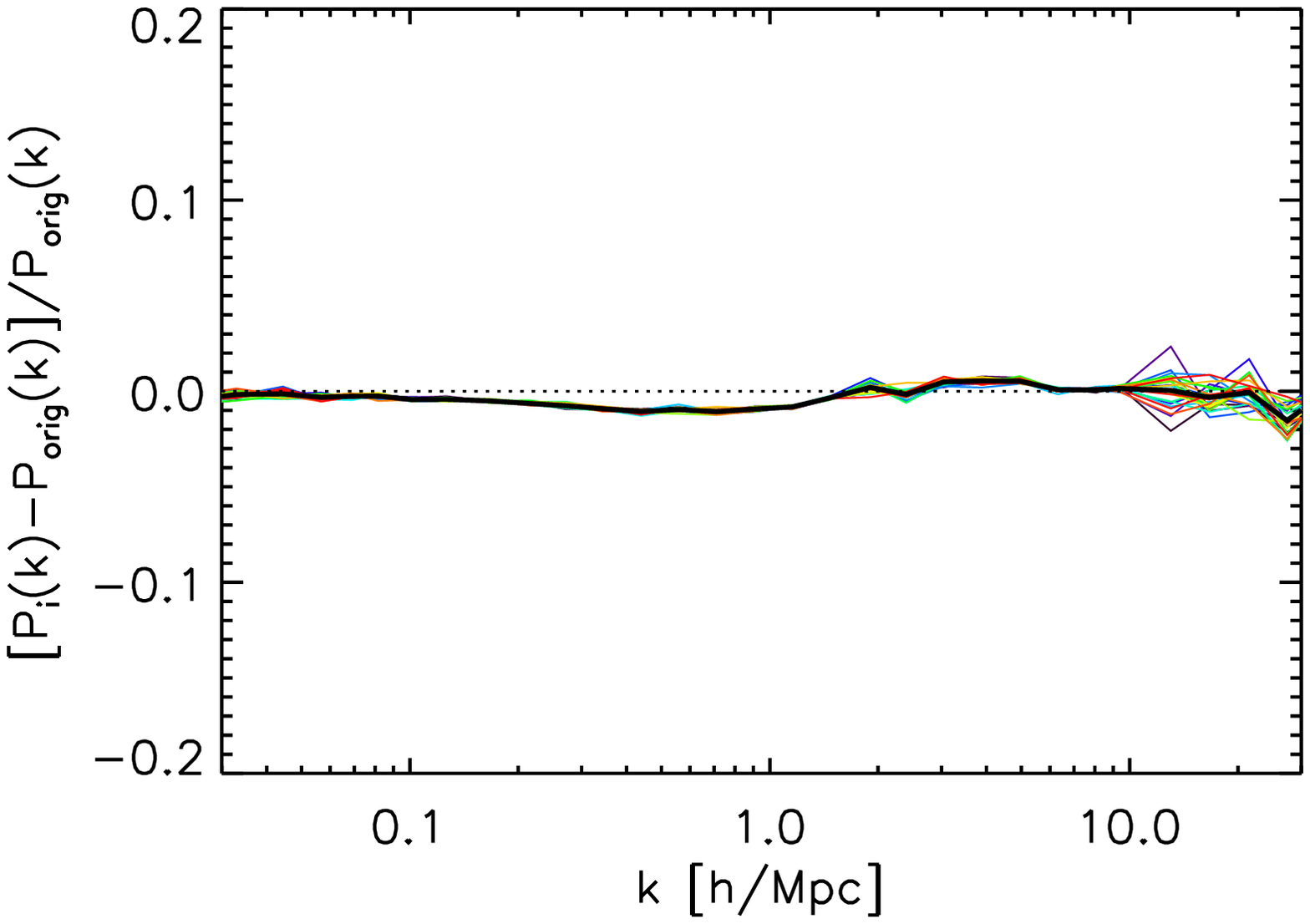} & &
\includegraphics[width=1.0\columnwidth, trim=7mm 8mm 0mm 8mm]{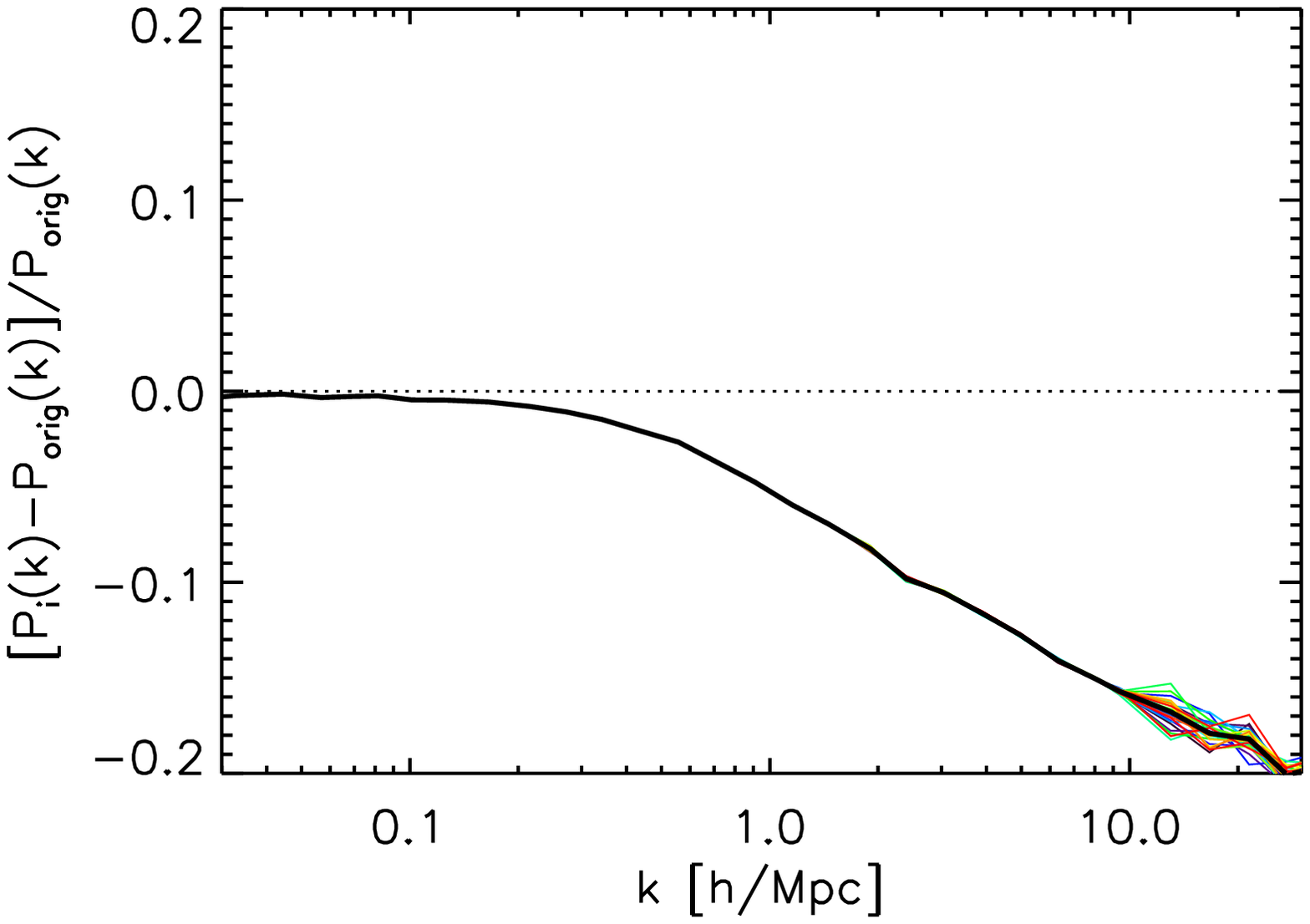}
\end{tabular}
\caption{Same as Figures \ref{correl_rot1} and \ref{correl_rot2}, but now for the fractional differences in the galaxy power spectrum versus the wave number $k$. \textit{Left:} Result when testing for alignment. Again a weak but systematic signal of a few per cent can be seen, now between $k \approx 0.1\kunit$ and $k \approx 2\kunit$. \textit{Right:} Result when testing for ellipticity. A monotonic decline in power sets in at $k \approx 0.1\kunit$, reaching roughly $20$ per cent at $k=30\kunit$, matching the result found in Figure~\ref{correl_rot2}. This again demonstrates the importance of taking the ellipsoidal shape of the galaxy distribution within haloes into account.}
\label{correl_rot3}
\end{center}
\end{figure*}

The ellipsoidal shape of the haloes, and the fact that the galaxy distribution follows this shape, is a more significant factor when modelling the galaxy distribution. Figure~\ref{correl_rot2} shows the result of applying independent rotations, which sphericalize the galaxy distributions within haloes. This substantially suppresses correlations for $r \la 2\runit$, with a $\sim 20$ per cent effect on the smallest scales probed here. The ellipsoidal shape of the galaxy distribution within haloes significantly reduces the typical separations of pairs within them. This is compensated by a $1-2$ per cent stronger correlation around $r \approx 3.5\runit$. We conclude that models that assume spherical profiles for the distribution of galaxies within haloes will underestimate the galaxy correlation function by up to $\sim 20$ per cent, depending on the smallest scale considered. Note that the scatter is even smaller than before on all scales, which is due to the increased number of degrees of freedom here. This result is in excellent agreement with that of \citet{Zu2008}, who applied similar methods to all galaxies from \citet{DeLuciaBlaizot2007} with $r$-band luminosities $M_r<-19$.

One might worry that the necessarily artificial treatment of orphan, or ``type 2'', galaxies in the galaxy formation models of G11 influences these results. G11 already showed that the inclusion of the orphans is critical if the radial distribution of galaxies within rich clusters in the Millennium Simulation is to agree both with observations and with the much higher-resolution MS-II. We have investigated effects on our analysis by repeating it with these galaxies removed, reducing our sample size by $\sim 24$ per cent and spoiling the relatively good agreement of its small-scale correlation with observation. This removal significantly amplifies the signal found for the effects of alignment. This is because the orphan galaxies are primarily located near halo centres. Once they are removed, galaxies that do contribute to the alignment signal receive more weight. Orphan removal also changes the signal found for the effects of ellipticity, modestly boosting it down to $r \approx 0.1\runit$. As a further check that the distribution of orphans in the simulation is realistic, we have examined the shapes of the galaxy distributions of massive haloes (specifically, $14<\log_{10}(M_\mathrm{FoF}/[\munit])<14.5$) in the MS and MS-II, again using G11's galaxy catalogues and considering only galaxies with stellar masses $M_*>10^9\munit$. The better mass resolution of the MS-II results in far fewer orphans in this mass range, and consequently the positions of galaxies in MS-II are determined more accurately. Nevertheless, the shapes of the galaxy distributions agree very well, thus implying that the distribution of orphans in the MS is consistent with the distribution of similar, but unstripped, galaxies in the MS-II. We also found that these shapes agree very well with those of the dark matter haloes themselves \citep[see][]{Bett2007}. A more detailed discussion of the shapes is beyond the scope of this paper.

For completeness, we have also compared the galaxy-galaxy power spectra of the rotated and unrotated samples. The results are shown in Figure~\ref{correl_rot3}. Three foldings in total were used to calculate the power spectra over the full range shown, each with a fold factor of six (i.e. each folding maps the particle distribution to $1/216$th of the volume). The power spectra were re-binned logarithmically to resemble the bins used for the galaxy correlation functions, and to reduce noise. Shot noise, which dominates the power for $k \ga 10\kunit$, was subtracted. The left-hand figure shows the fractional differences that result from applying bodily rotations. Just as for the correlation function, there is a weak but clear dip of $1-2$ per cent present which reflects the alignment of haloes with surrounding structure. As expected, the right-hand figure shows a much stronger decrease in power, up to $20$ per cent on the smallest scales considered. This again demonstrates the importance of taking the ellipsoidal distribution of galaxies within FoF groups into account in, for example, models that use that use the full shape of the power spectrum to extract cosmological parameters.

Our results differ from those found by \citet{SmithWatts2005}. Like ourselves, they find that the scale at which the contribution from alignments to the power spectrum is maximal is $\sim 0.5\kunit$, but they show that the relative contribution of alignments is strongly model-dependent, varying from $10^{-12}$ to $10$ per cent. Furthermore, they find that when haloes are assumed to be spherical, the power is higher than when they are ellipsoidal by up to $5$ per cent. Not only is the effect we find significantly stronger and increasing towards smaller scales up to at least $k=30\kunit$, but its sign is opposite. We attribute these differences to the fact that in both models explored by \citet{SmithWatts2005}, the radially averaged density profiles are not conserved when transforming the haloes from spherical to triaxial, making a comparison with our own results difficult.

\begin{figure}
\begin{center}
\includegraphics[width=1.0\columnwidth, trim=23mm 4mm 19mm 12mm]{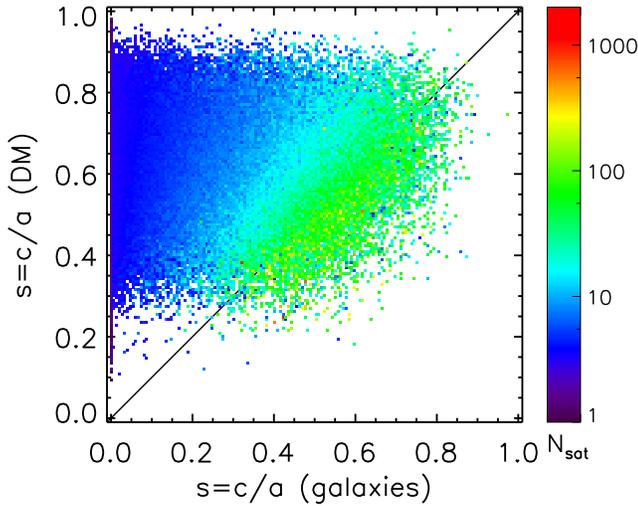}
\caption{A comparison of the halo shape measured from the dark matter and that measured from the galaxy distribution. Here shape is defined as the ratio of the smallest to the largest eigenvalue of the inertia tensor. The shape on the horizontal axis is computed using all galaxies satisfying $M_*>10^9\munit$ within $R_{200}$. For the shape on the vertical axis, only CDM particles belonging to the most-pronounced substructure are included, and only if the number of particles is at least 700. Pixels are colour-coded by mean number of galaxies per halo. The ``true'' shape is recovered more accurately when the number of galaxies increases; some scatter remains however, as the galaxy distribution does not perfectly trace the main subhalo.}
\label{ca-ca}
\end{center}
\end{figure}

\subsection{Shape-dependent galaxy bias}
\label{subsec:biasresults}
Having established that the shape of the galaxy distribution significantly affects the small-scale clustering, we now investigate how the shape-dependent assembly bias affects clustering. We first examine how well the shape measured from the galaxy distribution corresponds to that measured from the dark matter particles belonging to the most-pronounced substructure. Note that while the galaxies trace the shape of the FoF halo very well, this is not necessarily the case for its most-pronounced substructure. Additionally, even if the galaxy distribution traces the dark matter perfectly, the right shape may still not be recovered if only a small number of galaxies is available to sample the halo.

The results of this shape comparison are presented in Figure~\ref{ca-ca}. Here the shape measured from the galaxy distribution is shown on the horizontal axis, while the shape measured from the dark matter is on the vertical axis. Each pixel is colour-coded by the mean number of satellites, $N_\mathrm{sat}$, satisfying $M_*<10^9\munit$ and $|r_\mathrm{sat}-r_\mathrm{cen}|<R_{200}$. A low value of $s=c/a$ indicates that the halo is (measured to be) very aspherical, while a perfectly spherical halo would have $s=1$. It is immediately clear that a low number of satellites leads to a severe underestimate of $s$. This is expected: together with the central galaxy, any two satellites will define a plane, ensuring that $c=0$. It is only when the halo is sampled by a large enough ensemble of points that the inertia tensor can be determined accurately. Figure~\ref{ca-ca} illustrates that one needs $N_\mathrm{sat}\ga 30$ to get an unbiased and accurate shape estimate. However, there is always a significant amount of scatter around the diagonal. This is due to the galaxies tracing the mass, i.e. the shape of the whole FoF group, and not just that of the most-pronounced substructure.

\begin{figure}
\begin{center}
\includegraphics[width=1.0\columnwidth, trim=-3mm -13mm 4mm -3mm]{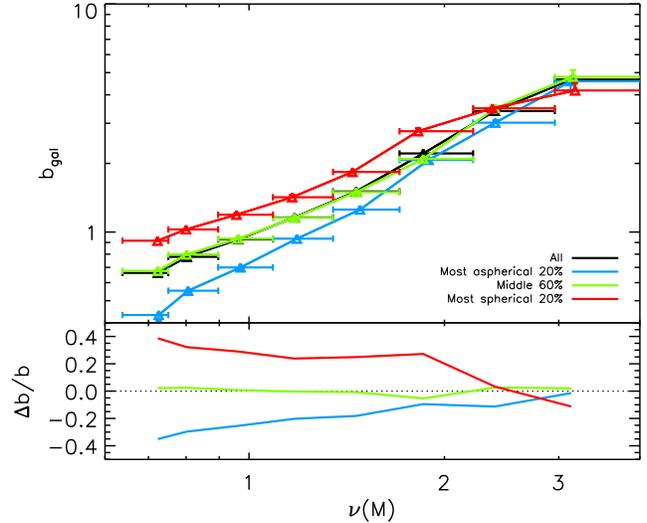}
\caption{The dependence of galaxy bias, $b_\mathrm{gal}$, on shape, as a function of peak height at $z=0$. The black line in the top panel shows the galaxy bias of the full galaxy sample, while coloured lines show the bias of subsamples split by halo shape measured from the dark matter distribution. In the bottom panel the fractional differences of the bias of these subsamples relative to the full sample are shown. The vertical error bars show $1\sigma$ deviations calculated from 50 bootstrap resamplings of the galaxy catalogues. We find that galaxies in the most spherical (aspherical) haloes are strongly biased (antibiased) relative to the full sample. The difference can be as much as $40$ per cent for $\nu \approx 0.7$. For $\nu \ga 2.5$ statistical uncertainties, due to the low number of high-mass haloes, begin to play an important role.}
\label{bias1}
\end{center}
\end{figure}

Next, we split our galaxy sample by the shape measured from the halo dark matter distribution and calculate the galaxy bias factor as function of equivalent peak height, $b_\mathrm{gal}(\nu)$, following the method described in \S\ref{subsec:bias}. As we can only use galaxies for which a dark matter shape has been determined -- i.e. those in FoF haloes of which the most massive substructure is comprised of at least 700 particles -- our galaxy sample is reduced to $2\,953\,050$ galaxies. The results are shown in Figure~\ref{bias1}. Here the upper panel shows the galaxy bias determined for each sample at a given equivalent peak height. The points show the median value, horizontal error bars indicate the width of the bin, and vertical error bars show $1\sigma$ deviations calculated from 50 bootstrap resamplings of the galaxy catalogues. The black line shows $b_\mathrm{gal}(\nu)$ for the full sample of galaxies, while the coloured lines show the bias for the different subsamples. In the bottom panel, the fractional difference between these subsamples and the full sample is shown. It is immediately clear that there is a strong dependence on shape: at the lowest equivalent peak heights probed here, the bias of the galaxies in the most spherical (aspherical) 20 per cent of haloes is up to 40 per cent higher (lower) than that of the full sample. This shows that the shape-dependence of the halo bias found by \citet{Bett2007} and \citet{FaltenbacherWhite2010} is also strongly present in the clustering of the galaxies. The effect grows weaker with increasing $\nu$. However, for $\nu \ga 2.5$ statistical uncertainties begin to dominate, due to the low number of haloes available at high equivalent peak heights. Note that this is not seen in the data of \citet{FaltenbacherWhite2010} as they combine the data from different redshifts, while we only consider $z=0$.

\begin{figure}
\begin{center}
\includegraphics[width=1.0\columnwidth, trim=-3mm -13mm 4mm -3mm]{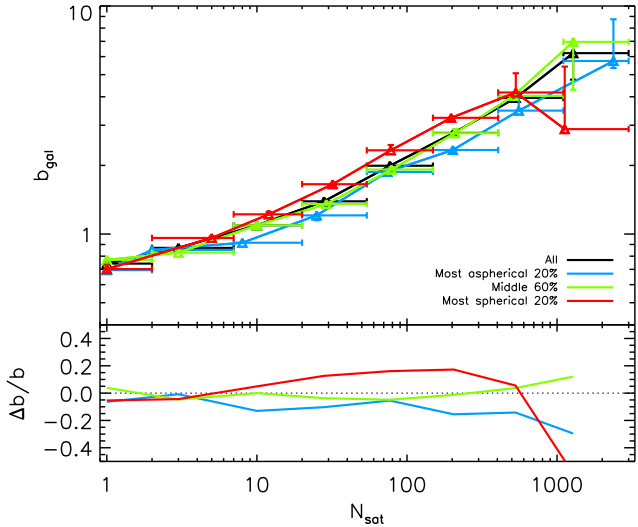}
\caption{As Figure \ref{bias1}, but now showing the bias as a function of the number of satellite galaxies and split by the halo shape measured from galaxies within $R_{200}$. At low $N_\mathrm{sat}$ no significant shape-dependence of $b_\mathrm{gal}$ is recovered, due to the extremely unreliable shape determinations that follow from using only a handful of galaxies to sample the halo. For $10 \la N_\mathrm{sat} \la 400$, however, we find again that galaxies in more spherical (aspherical) haloes have a significantly higher (lower) bias than average. At higher $N_\mathrm{sat}$ our results are again dominated by poor statistics.}
\label{bias2}
\end{center}
\end{figure}

We then repeat this exercise, but this time we split our galaxy sample by the shape measured from the distribution of the galaxies themselves. However, we expect low-mass haloes to host only a few satellite galaxies, leading to very unreliable estimates of the halo shape (see Figure~\ref{ca-ca}). In order to separate the signal we are looking for -- i.e. the shape-dependence of the galaxy bias  -- from the unwanted bias introduced by using too few galaxies in the shape measurement, we now consider $b_\mathrm{gal}$ as a function of the number of satellites per halo, $N_\mathrm{sat}$, instead of the equivalent peak height $\nu$. An additional advantage of this approach is that $N_\mathrm{sat}$ is directly observable. We note, however, that we obtain almost identical results when considering galaxy bias as a function of peak height instead of $N_\mathrm{sat}$. Since we do not use the dark matter particle data in this case, we are no longer constrained by needing haloes with at least 700 particles. However, as we can now only consider haloes with at least one satellite galaxy that satisfies both $M_*>10^9\munit$ and $|r_\mathrm{sat}-r_\mathrm{cen}|<R_{200}$, we are left with a sample of $2\,566\,441$ galaxies. The results are shown in Figure~\ref{bias2}. At the lowest value of $N_\mathrm{sat}$, no significant effect can be seen. But as $N_\mathrm{sat}$ grows, increasing the accuracy of the shape determinations, we again see a clear dependence of $b_\mathrm{gal}$ on the shape $s$: galaxies in more spherical haloes are $\sim 20$ per cent more strongly clustered than average. The inverse is true for galaxies in the most aspherical haloes. When $N_\mathrm{sat}$ grows too high our results are once more dominated by statistical errors, due to the low number of high-mass haloes (hosting at least several hundreds of satellites) available.

These results show that halo assembly bias in the form of a shape-dependent clustering strength propagates to the clustering of galaxies, and can therefore in principle be measured in sufficiently large surveys. In order to carry out such a task, a large galaxy survey with appropriately defined group catalogues is needed.

\section{Summary}
\label{sec:summary}
We have investigated the effects of halo alignment with larger scale structure and of halo ellipticity on galaxy correlation functions, using the Millennium Simulations \citep{Springel2005,Boylan-Kolchin2009} and the galaxy formation models of \citet{Guo2011}. By rotating satellite galaxies in FoF groups around their central galaxies, either coherently for each halo or independently for each satellite, and then comparing the correlation function of the resulting galaxy distribution to the original one, we were able to quantify the importance of taking halo alignment and non-sphericity into account. Furthermore, by measuring the shape of the haloes as traced by the galaxies we were able to investigate the propagation of shape-dependent assembly bias to the clustering of galaxies. Only galaxies with stellar masses $M_*>10^9\munit$ were considered in our analysis, though we note that increasing this mass limit by a factor of ten does not influence our results. Our findings can be summarized as follows:
\begin{itemize}
\item The effects on the galaxy correlation function of the alignment of haloes with larger-scale structure are small. The main effect of disrupting this alignment is a $2$ per cent reduction in correlation amplitude around $r \approx 1.8\runit$, with minor effects of at most $1$ per cent at smaller scales.
\item The ellipsoidal shapes of the galaxy distributions within individual haloes have a much stronger influence on galaxy correlations. By sphericalizing these galaxy distributions (i.e. randomizing the angular positions of satellites while keeping the distance from the central galaxy fixed), the correlation function is raised by up to $2$ per cent around $r \approx 3.5\runit$, but greatly reduced for $r \la 1.5\runit$, by up to $\sim 20$ per cent on the smallest scale probed, $r=30\runitk$. This confirms the results of \citet{Zu2008}. The effect on the galaxy power spectrum extends to scales as large as $k=0.1\kunit$.
\item The assembly bias of haloes, as characterized by the dependence of clustering on halo shape, is reflected in the clustering of galaxies. The effect is strongest at low equivalent peak heights: at $\nu \approx 0.7$, the galaxy bias of galaxies in the 20 per cent most spherical and most aspherical haloes deviate from the average by 40 per cent.
\item Even if the shape of the halo cannot be measured directly, but is instead estimated from galaxies within one virial radius of the central galaxy, the effect of assembly bias is clearly visible.
\end{itemize}
By using the plane-parallel approximation and ignoring evolution, we have checked that comparable results are obtained for the effects of alignment and ellipticity on the redshift-space correlation functions. Models that assume a spherically symmetric profile for the galaxy distribution, such as HOD models, will therefore significantly underestimate galaxy correlations and power spectra on sub-Mpc scales.

Furthermore, we have demonstrated that the shapes of haloes and of the galaxy distributions within them can be strongly correlated with their clustering. This effect should be measurable in galaxy redshift surveys. With the help of extensive group catalogues it should therefore be possible to measure assembly bias directly.

\section*{Acknowledgments}
The authors thank Andreas Faltenbacher for kindly providing dark matter shape determinations for haloes in the Millennium Simulation. The Millennium Simulation databases used in this paper and the web application providing online access to them were constructed as part of the activities of the German Astrophysical Virtual Observatory. This work was supported by the Marie Curie Initial Training Network CosmoComp (PITN-GA-2009-238356) and by Advanced Grant 246797 "GALFORMOD" from the European Research Council.

\bibliographystyle{mn2e}
\bibliography{mybib}

\begin{thebibliography}{53}
\expandafter\ifx\csname natexlab\endcsname\relax\def\natexlab#1{#1}\fi

\bibitem[{{Allgood} {et~al}\mbox{.}(2006){Allgood}, {Flores}, {Primack},
  {Kravtsov}, {Wechsler}, {Faltenbacher}, \& {Bullock}}]{Allgood2006}
{Allgood} B., {Flores} R.~A., {Primack} J.~R., {Kravtsov} A.~V., {Wechsler}
  R.~H., {Faltenbacher} A., {Bullock} J.~S., 2006, \mnras, 367, 1781

\bibitem[{{Angulo} {et~al}\mbox{.}(2009){Angulo}, {Lacey}, {Baugh}, \&
  {Frenk}}]{Angulo2009}
{Angulo} R.~E., {Lacey} C.~G., {Baugh} C.~M., {Frenk} C.~S., 2009, \mnras, 399,
  983

\bibitem[{{Angulo} \& {White}(2010)}]{AnguloWhite2010}
{Angulo} R.~E., {White} S.~D.~M., 2010, \mnras, 405, 143

\bibitem[{{Bailin} \& {Steinmetz}(2005)}]{Bailin2005}
{Bailin} J., {Steinmetz} M., 2005, \apj, 627, 647

\bibitem[{{Barriga} \& {Gazta{\~n}aga}(2002)}]{Barriga2002}
{Barriga} J., {Gazta{\~n}aga} E., 2002, \mnras, 333, 443

\bibitem[{{Baugh}(2006)}]{Baugh2006}
{Baugh} C.~M., 2006, Reports on Progress in Physics, 69, 3101

\bibitem[{{Bett} {et~al}\mbox{.}(2007){Bett}, {Eke}, {Frenk}, {Jenkins},
  {Helly}, \& {Navarro}}]{Bett2007}
{Bett} P., {Eke} V., {Frenk} C.~S., {Jenkins} A., {Helly} J., {Navarro} J.,
  2007, \mnras, 376, 215

\bibitem[{{Binggeli}(1982)}]{Binggeli1982}
{Binggeli} B., 1982, \aap, 107, 338

\bibitem[{{Boylan-Kolchin}, {Ma} \& {Quataert}(2008){Boylan-Kolchin}, {Ma}, \&
  {Quataert}}]{Boylan-Kolchin2008}
{Boylan-Kolchin} M., {Ma} C.-P., {Quataert} E., 2008, \mnras, 383, 93

\bibitem[{{Boylan-Kolchin} {et~al}\mbox{.}(2009){Boylan-Kolchin}, {Springel},
  {White}, {Jenkins}, \& {Lemson}}]{Boylan-Kolchin2009}
{Boylan-Kolchin} M., {Springel} V., {White} S.~D.~M., {Jenkins} A., {Lemson}
  G., 2009, \mnras, 398, 1150

\bibitem[{{Carter} \& {Metcalfe}(1980)}]{CarterMetcalfe1980}
{Carter} D., {Metcalfe} N., 1980, \mnras, 191, 325

\bibitem[{{Colberg} {et~al}\mbox{.}(1999){Colberg}, {White}, {Jenkins}, \&
  {Pearce}}]{Colberg1999}
{Colberg} J.~M., {White} S.~D.~M., {Jenkins} A., {Pearce} F.~R., 1999, \mnras,
  308, 593

\bibitem[{{Cole} {et~al}\mbox{.}(1994){Cole}, {Aragon-Salamanca}, {Frenk},
  {Navarro}, \& {Zepf}}]{Cole1994}
{Cole} S., {Aragon-Salamanca} A., {Frenk} C.~S., {Navarro} J.~F., {Zepf} S.~E.,
  1994, \mnras, 271, 781

\bibitem[{{Colless} {et~al}\mbox{.}(2001){Colless}, {Dalton}, {Maddox},
  {Sutherland}, {Norberg}, {Cole}, {Bland-Hawthorn}, {Bridges}, {Cannon},
  {Collins}, {Couch}, {Cross}, {Deeley}, {De Propris}, {Driver}, {Efstathiou},
  {Ellis}, {Frenk}, {Glazebrook}, {Jackson}, {Lahav}, {Lewis}, {Lumsden},
  {Madgwick}, {Peacock}, {Peterson}, {Price}, {Seaborne}, \&
  {Taylor}}]{Colless2001}
{Colless} M. {et~al.}, 2001, \mnras, 328, 1039

\bibitem[{{Cooray} \& {Sheth}(2002)}]{CooraySheth2002}
{Cooray} A., {Sheth} R., 2002, \physrep, 372, 1

\bibitem[{{Croton}, {Gao} \& {White}(2007){Croton}, {Gao}, \&
  {White}}]{Croton2007}
{Croton} D.~J., {Gao} L., {White} S.~D.~M., 2007, \mnras, 374, 1303

\bibitem[{{Davis} {et~al}\mbox{.}(1985){Davis}, {Efstathiou}, {Frenk}, \&
  {White}}]{Davis1985}
{Davis} M., {Efstathiou} G., {Frenk} C.~S., {White} S.~D.~M., 1985, \apj, 292,
  371

\bibitem[{{De Lucia} \& {Blaizot}(2007)}]{DeLuciaBlaizot2007}
{De Lucia} G., {Blaizot} J., 2007, \mnras, 375, 2

\bibitem[{{Eriksen} {et~al}\mbox{.}(2004){Eriksen}, {Lilje}, {Banday}, \&
  {G{\'o}rski}}]{Eriksen2004}
{Eriksen} H.~K., {Lilje} P.~B., {Banday} A.~J., {G{\'o}rski} K.~M., 2004,
  \apjs, 151, 1

\bibitem[{{Faltenbacher} \& {White}(2010)}]{FaltenbacherWhite2010}
{Faltenbacher} A., {White} S.~D.~M., 2010, \apj, 708, 469

\bibitem[{{Gao}, {Springel} \& {White}(2005){Gao}, {Springel}, \&
  {White}}]{Gao2005}
{Gao} L., {Springel} V., {White} S.~D.~M., 2005, \mnras, 363, L66

\bibitem[{{Gao} \& {White}(2007)}]{GaoWhite2007}
{Gao} L., {White} S.~D.~M., 2007, \mnras, 377, L5

\bibitem[{{Gil-Mar{\'{\i}}n}, {Jimenez} \& {Verde}(2011){Gil-Mar{\'{\i}}n},
  {Jimenez}, \& {Verde}}]{Gil-Marin2011}
{Gil-Mar{\'{\i}}n} H., {Jimenez} R., {Verde} L., 2011, \mnras, 414, 1207

\bibitem[{{Giocoli} {et~al}\mbox{.}(2010){Giocoli}, {Bartelmann}, {Sheth}, \&
  {Cacciato}}]{Giocoli2010}
{Giocoli} C., {Bartelmann} M., {Sheth} R.~K., {Cacciato} M., 2010, \mnras, 408,
  300

\bibitem[{{Guo} {et~al}\mbox{.}(2012){Guo}, {White}, {Angulo}, {Henriques},
  {Lemson}, {Boylan-Kolchin}, {Thomas}, \& {Short}}]{Guo2012}
{Guo} Q., {White} S., {Angulo} R.~E., {Henriques} B., {Lemson} G.,
  {Boylan-Kolchin} M., {Thomas} P., {Short} C., 2012, preprint (arXiv:1206.0052)

\bibitem[{{Guo} {et~al}\mbox{.}(2011){Guo}, {White}, {Boylan-Kolchin}, {De
  Lucia}, {Kauffmann}, {Lemson}, {Li}, {Springel}, \& {Weinmann}}]{Guo2011}
{Guo} Q. {et~al.}, 2011, \mnras, 413, 101

\bibitem[{{Jenkins} {et~al}\mbox{.}(1998){Jenkins}, {Frenk}, {Pearce},
  {Thomas}, {Colberg}, {White}, {Couchman}, {Peacock}, {Efstathiou}, \&
  {Nelson}}]{Jenkins1998}
{Jenkins} A. {et~al.}, 1998, \apj, 499, 20

\bibitem[{{Jing} \& {Suto}(2002)}]{JingSuto2002}
{Jing} Y.~P., {Suto} Y., 2002, \apj, 574, 538

\bibitem[{{Jing}, {Suto} \& {Mo}(2007){Jing}, {Suto}, \& {Mo}}]{Jing2007}
{Jing} Y.~P., {Suto} Y., {Mo} H.~J., 2007, \apj, 657, 664

\bibitem[{{Kauffmann} {et~al}\mbox{.}(1999){Kauffmann}, {Colberg}, {Diaferio},
  \& {White}}]{Kauffmann1999}
{Kauffmann} G., {Colberg} J.~M., {Diaferio} A., {White} S.~D.~M., 1999, \mnras,
  303, 188

\bibitem[{{Kauffmann}, {White} \& {Guiderdoni}(1993){Kauffmann}, {White}, \&
  {Guiderdoni}}]{Kauffmann1993}
{Kauffmann} G., {White} S.~D.~M., {Guiderdoni} B., 1993, \mnras, 264, 201

\bibitem[{{Komatsu} {et~al}\mbox{.}(2011){Komatsu}, {Smith}, {Dunkley},
  {Bennett}, {Gold}, {Hinshaw}, {Jarosik}, {Larson}, {Nolta}, {Page},
  {Spergel}, {Halpern}, {Hill}, {Kogut}, {Limon}, {Meyer}, {Odegard}, {Tucker},
  {Weiland}, {Wollack}, \& {Wright}}]{Komatsu2011}
{Komatsu} E. {et~al.}, 2011, \apjs, 192, 18

\bibitem[{{Kravtsov} {et~al}\mbox{.}(2004){Kravtsov}, {Berlind}, {Wechsler},
  {Klypin}, {Gottl{\"o}ber}, {Allgood}, \& {Primack}}]{Kravtsov2004}
{Kravtsov} A.~V., {Berlind} A.~A., {Wechsler} R.~H., {Klypin} A.~A.,
  {Gottl{\"o}ber} S., {Allgood} B., {Primack} J.~R., 2004, \apj, 609, 35

\bibitem[{{Navarro}, {Frenk} \& {White}(1997){Navarro}, {Frenk}, \&
  {White}}]{Navarro1997}
{Navarro} J.~F., {Frenk} C.~S., {White} S.~D.~M., 1997, \apj, 490, 493

\bibitem[{{Paz} {et~al}\mbox{.}(2011){Paz}, {Sgr{\'o}}, {Merch{\'a}n}, \&
  {Padilla}}]{Paz2011}
{Paz} D.~J., {Sgr{\'o}} M.~A., {Merch{\'a}n} M., {Padilla} N., 2011, \mnras,
  414, 2029

\bibitem[{{Ragone-Figueroa} {et~al}\mbox{.}(2010){Ragone-Figueroa}, {Plionis},
  {Merch{\'a}n}, {Gottl{\"o}ber}, \& {Yepes}}]{Ragone-Figueroa2010}
{Ragone-Figueroa} C., {Plionis} M., {Merch{\'a}n} M., {Gottl{\"o}ber} S.,
  {Yepes} G., 2010, \mnras, 407, 581

\bibitem[{{S{\'a}nchez}, {Baugh} \& {Angulo}(2008){S{\'a}nchez}, {Baugh}, \&
  {Angulo}}]{Sanchez2008}
{S{\'a}nchez} A.~G., {Baugh} C.~M., {Angulo} R., 2008, \mnras, 390, 1470

\bibitem[{{Smargon} {et~al}\mbox{.}(2012){Smargon}, {Mandelbaum}, {Bahcall}, \&
  {Niederste-Ostholt}}]{Smargon2012}
{Smargon} A., {Mandelbaum} R., {Bahcall} N., {Niederste-Ostholt} M., 2012,
  \mnras, 423, 856

\bibitem[{{Smith} \& {Watts}(2005)}]{SmithWatts2005}
{Smith} R.~E., {Watts} P.~I.~R., 2005, \mnras, 360, 203

\bibitem[{{Spergel} {et~al}\mbox{.}(2003){Spergel}, {Verde}, {Peiris},
  {Komatsu}, {Nolta}, {Bennett}, {Halpern}, {Hinshaw}, {Jarosik}, {Kogut},
  {Limon}, {Meyer}, {Page}, {Tucker}, {Weiland}, {Wollack}, \&
  {Wright}}]{Spergel2003}
{Spergel} D.~N. {et~al.}, 2003, \apjs, 148, 175

\bibitem[{{Splinter} {et~al}\mbox{.}(1997){Splinter}, {Melott}, {Linn}, {Buck},
  \& {Tinker}}]{Splinter1997}
{Splinter} R.~J., {Melott} A.~L., {Linn} A.~M., {Buck} C., {Tinker} J., 1997,
  \apj, 479, 632

\bibitem[{{Springel}(2010)}]{Springel2010}
{Springel} V., 2010, \araa, 48, 391

\bibitem[{{Springel} {et~al}\mbox{.}(2005){Springel}, {White}, {Jenkins},
  {Frenk}, {Yoshida}, {Gao}, {Navarro}, {Thacker}, {Croton}, {Helly},
  {Peacock}, {Cole}, {Thomas}, {Couchman}, {Evrard}, {Colberg}, \&
  {Pearce}}]{Springel2005}
{Springel} V. {et~al.}, 2005, \nat, 435, 629

\bibitem[{{Springel} {et~al}\mbox{.}(2001){Springel}, {White}, {Tormen}, \&
  {Kauffmann}}]{Springel2001}
{Springel} V., {White} S.~D.~M., {Tormen} G., {Kauffmann} G., 2001, \mnras,
  328, 726

\bibitem[{{Tormen}, {Bouchet} \& {White}(1997){Tormen}, {Bouchet}, \&
  {White}}]{Tormen1997}
{Tormen} G., {Bouchet} F.~R., {White} S.~D.~M., 1997, \mnras, 286, 865

\bibitem[{{Vera-Ciro} {et~al}\mbox{.}(2011){Vera-Ciro}, {Sales}, {Helmi},
  {Frenk}, {Navarro}, {Springel}, {Vogelsberger}, \& {White}}]{Vera-Ciro2011}
{Vera-Ciro} C.~A., {Sales} L.~V., {Helmi} A., {Frenk} C.~S., {Navarro} J.~F.,
  {Springel} V., {Vogelsberger} M., {White} S.~D.~M., 2011, \mnras, 416, 1377

\bibitem[{{Wechsler} {et~al}\mbox{.}(2006){Wechsler}, {Zentner}, {Bullock},
  {Kravtsov}, \& {Allgood}}]{Wechsler2006}
{Wechsler} R.~H., {Zentner} A.~R., {Bullock} J.~S., {Kravtsov} A.~V., {Allgood}
  B., 2006, \apj, 652, 71

\bibitem[{{West}(1989)}]{West1989}
{West} M.~J., 1989, \apj, 347, 610

\bibitem[{{Wetzel} {et~al}\mbox{.}(2007){Wetzel}, {Cohn}, {White}, {Holz}, \&
  {Warren}}]{Wetzel2007}
{Wetzel} A.~R., {Cohn} J.~D., {White} M., {Holz} D.~E., {Warren} M.~S., 2007,
  \apj, 656, 139

\bibitem[{{White} \& {Frenk}(1991)}]{WhiteFrenk1991}
{White} S.~D.~M., {Frenk} C.~S., 1991, \apj, 379, 52

\bibitem[{{Zheng} {et~al}\mbox{.}(2005){Zheng}, {Berlind}, {Weinberg},
  {Benson}, {Baugh}, {Cole}, {Dav{\'e}}, {Frenk}, {Katz}, \&
  {Lacey}}]{Zheng2005}
{Zheng} Z. {et~al.}, 2005, \apj, 633, 791

\bibitem[{{Zhu} {et~al}\mbox{.}(2006){Zhu}, {Zheng}, {Lin}, {Jing}, {Kang}, \&
  {Gao}}]{Zhu2006}
{Zhu} G., {Zheng} Z., {Lin} W.~P., {Jing} Y.~P., {Kang} X., {Gao} L., 2006,
  \apjl, 639, L5

\bibitem[{{Zu} {et~al}\mbox{.}(2008){Zu}, {Zheng}, {Zhu}, \& {Jing}}]{Zu2008}
{Zu} Y., {Zheng} Z., {Zhu} G., {Jing} Y.~P., 2008, \apj, 686, 41

\end{thebibliography}

\bsp
\label{lastpage}
\end{document}